\documentclass[a4paper,fleqn]{cas-sc}

\usepackage[authoryear,longnamesfirst]{natbib}

\usepackage[nolist]{acronym}
\usepackage[capitalise]{cleveref}
\usepackage{placeins}
\usepackage{color}

\definecolor{IMV1}{rgb}{0.64, 0.0, 0.0}

\def\tsc#1{\csdef{#1}{\textsc{\lowercase{#1}}\xspace}}
\tsc{WGM}
\tsc{QE}
\tsc{EP}
\tsc{PMS}
\tsc{BEC}
\tsc{DE}

\begin{document}
\let\WriteBookmarks\relax
\def\floatpagepagefraction{1}
\def\textpagefraction{.001}

\shorttitle{Controlling Vortices through Tip Permeability}

\shortauthors{Yabin Liu et~al.}

\title [mode = title]{Controlling Tip Vortices and Cavitation through Tip Permeability for Tidal Turbines}                  


\author[1,2]{Yabin Liu}[type=editor,
                        orcid=0000-0002-0150-5671]
\credit{Conceived, designed, and supervised the project; led the development of the numerical model and the data analysis; wrote the first draft of the manuscript.}
\cormark[1]
\ead{yl742@cam.ac.uk}
\ead[url]{https://www.eng.cam.ac.uk/profiles/yl742}

\author[1,2]{Junchen Tan}
\credit{Prepared and performed the experiments, post-processed the experimental data, and edited the manuscript.}

\author[3]{Richard H.~J. Willden}
\credit{Co-supervised the project, advised on the analysis of the results, and edited the manuscript.}

\author[1]{Paul Gary Tucker}
\credit{Co-supervised the project, advised on the simulations, and edited the manuscript.}

\author[2]{Ignazio Maria Viola}
\credit{Co-supervised the project, advised on the experiments and simulations, and edited the manuscript.}

\cortext[cor1]{Corresponding author}

\address[1]{Department of Engineering, University of Cambridge, Cambridge, CB2 1PZ, UK}
\address[2]{School of Engineering, University of Edinburgh, Edinburgh, EH9 3BF, UK}
\address[3]{Department of Engineering Science, University of Oxford, Oxford, OX1 3PJ, UK}

\begin{abstract}
Blade-tip vortices can lead to wakes, cavitation and noise, and their control remains a significant challenge for tidal and wind turbines. In the present work, we propose controlling tip vortices through local permeability on a model-scale horizontal-axis turbine. The numerical investigation follows a rigorous validation and verification process. The tip permeability is modelled by including a porous zone over the blade tip, within which Darcy's law is applied. The results demonstrate that there is an optimal range of permeability, corresponding to a non-dimensional Darcy number, $Da$, of around $10^{-5}$, that can substantially decrease the tip vortex intensity. The revealed flow physics show that the permeable tip can effectively enlarge the vortex viscous core radius with little change to the vortex circulation. The permeable tip treatment can increase the minimal pressure-coefficient at the vortex core by up to 63\%, which significantly alleviates the cavitation risk. This approach has negligible influence on the turbine's energy-harvesting performance because the spanwise extent of the permeable zone is only in the order of 0.1\% turbine diameter. Our findings demonstrate this approach's great promise to break the upper tip-speed ratio limit capped by cavitation for tidal turbines, contributing to developing more efficient and resilient turbines.
\end{abstract}


\begin{highlights}


\item A novel approach of controlling tip vortices through permeable tip treatment is proposed and investigated. 

\item There is an optimal range of permeability that can suppress the pressure-drop with the tip vortices by up to 63\%.

\item Permeable tip treatment shows great potential for breaking the upper tip-speed ratio limit capped by cavitation.

\item The vortex intensity is mitigated because of the enhanced vortex diffusion and enlarged vortex dimension.
\end{highlights}


\maketitle

\begin{acronym}
    \acro{cfd}[CFD]{Computational Fluid Dynamics}
    \acro{tlv}[TLV]{tip leakage vortex}
    \acro{ptv}[PTV]{primary tip vortex}
    \acro{tsr}[TSR]{tip speed ratio}
    \acro{ptlv}[PTLV]{primary tip leakage vortex}
    \acro{siv}[SIV]{secondary induced vortex}
    \acro{tsv}[TSV]{tip separation vortex}
    \acro{spod}[SPOD]{Spectral Proper Orthogonal Decomposition}
    \acro{ps}[PS]{pressure side}
    \acro{ss}[SS]{suction side}
    \acro{piv}[PIV]{Particle Image Velocimetry}
    \acro{rans}[RANS]{Reynolds-averaged Navier-Stokes}
    \acro{les}[LES]{Large Eddy Simulation}
    \acro{zles}[ZLES]{Zonalised \acs{les}}
    \acro{des}[DES]{Detached Eddy Simulation}
    \acro{ddes}[DDES]{Delayed Detached Eddy Simulation}
    \acro{iddes}[IDDES]{Improved Delayed Detached Eddy Simulation}
    \acro{tke}[TKE]{turbulent kinetic energy}
\end{acronym}

\section{Introduction}

\begin{figure}
	\centering
		\includegraphics[scale=.6]{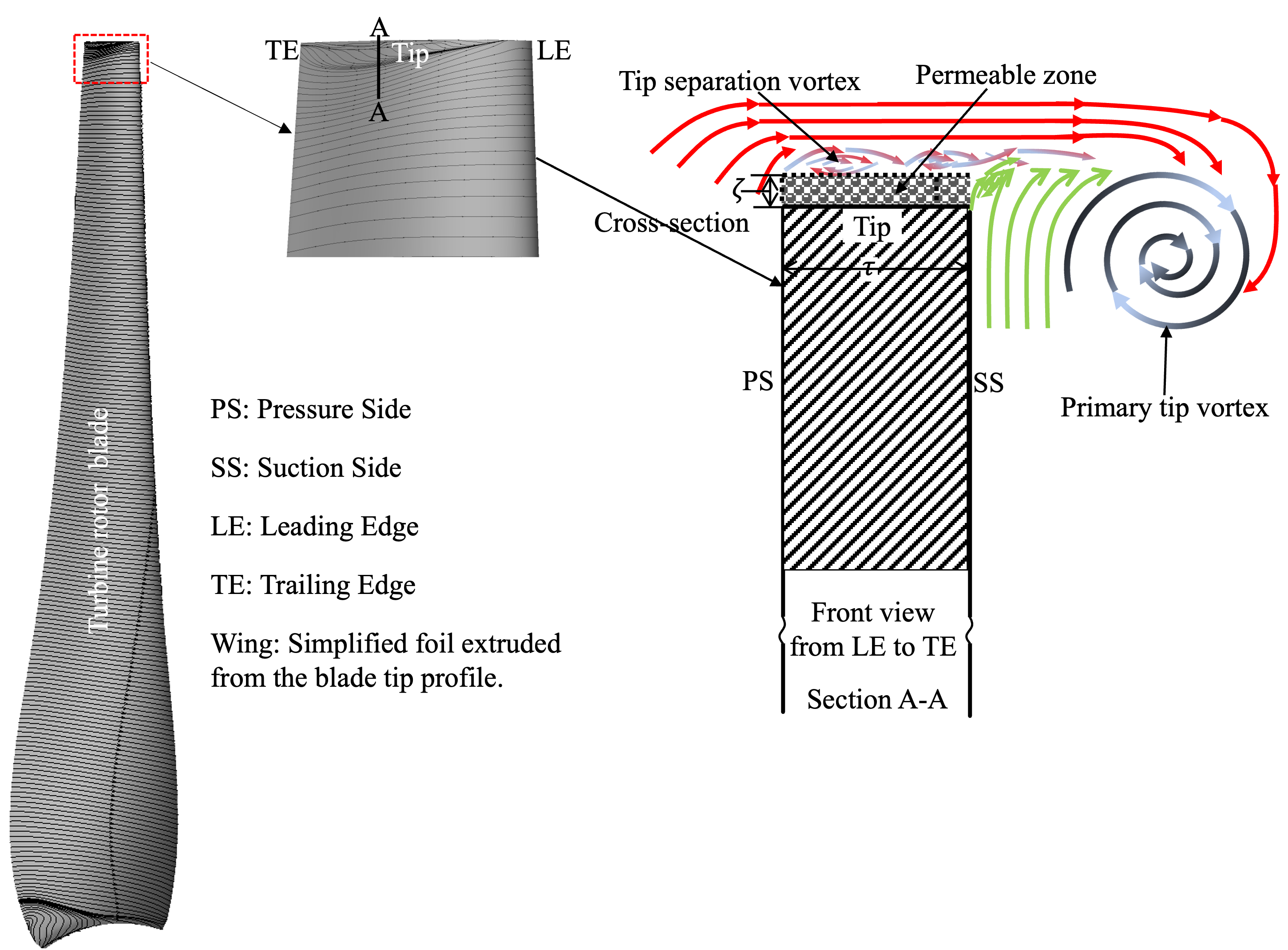}
	\caption{Schematic of tip vortices and permeable tip treatment on a tidal turbine blade. $\zeta$ denotes the spanwise scope of the permeable zone;  $\tau$ denotes the blade tip thickness.}
	\label{fig_schematic of permeable tip}
\end{figure}

Harnessing the power of renewable sources, such as wind and oceans, stands as a crucial stride in our world's transition to net zero. \cite{gwec2024} states that wind energy now meets around 10\% of global electricity demand, a milestone driven by substantial contributions from countries like China, which alone added 75 GW, accounting for two-thirds of the new capacity worldwide in 2023. In addition, tidal power shows great potential to serve as a significant part of the renewable energy scheme in future, meeting 11\% of annual electricity demands in the UK (\cite{coles2021review}). However, wind and tidal turbines, which are critical for converting renewable energy into electricity, face significant challenges related to blade-tip vortices. Tip vortices arise from the pressure difference across the wing or blade tip, creating a trailing swirl, shown in \Cref{fig_schematic of permeable tip}. These vortices, along with the associated wake (\cite{lignarolo2015tip, posa2021instability}), cavitation (\cite {liu2018review, wimshurst2018cavitation, wang2022modal}), and noise (\cite{ramachandran2014wind, bai2022investigation}) problems, can greatly impact the efficiency and resilience of wing- and blade-based systems, such as underwater vehicles and turbine rotors. In particular, tidal turbines operate underwater and their blade tips experience the highest flow speed, and thus cavitation occurs due to low pressure inside the tip vortices at a high \ac{tsr}. Moreover, the risk of blade erosion and damage is high due to the corrosive nature of seawater (\cite{hou2020cavitation}). These problems constrain the further increase of the turbines' \ac{tsr}, which is beneficial for both aero-hydrodynamic efficiency and reducing the size of powertrain components (\cite{ning2014understanding, wimshurst2018cavitation}).

In turbomachinery, two primary types of tip flows can occur: tip leakage vortices, when there is a gap between a rotating blade tip and casing wall, and tip vortices when there is no casing wall. Despite their differences, the underlying physics of these vortices share remarkable similarities. The intensity of tip (leakage) vortices, which is closely related to the local pressure-drop around the vortex core (\cite{ye2023dynamics, dreyer2014mind}) and the vortex breakdown (\cite{posa2024influence,liu2024tip}), has a dominant influence on the cavitation risks, flow unsteadiness, and wake recovery. To suppress the \ac{ptlv}, we introduced local micro-jets passively generated by a groove at the tip to interfere with the \ac{ptlv}, and this approach shows a significant potential to suppress the \ac{ptlv} and associated cavitation (\cite{liu2020influence, liu2018method, han2022method}). However, the experimental work followed by \cite{jiang2022groove} shows that the shrinking groove added on the tip may generate new cavitation because of the high local flow speed inside the groove itself. This suggests that the groove design needs to be improved by avoiding centralising the jet flow inside a single groove.

Bio-inspired research in natural systems, such as the flight of dandelion seeds, has demonstrated the critical role of permeability and porosity in controlling vortical structures (\cite{cummins2018separated, bose2023porous}). Inspired by the above, we propose a novel approach to controlling tip vortices with permeable tip treatment, aiming to mitigate the intensity of tip vortices and the associated cavitation risks. The tip permeability, modelled by a confined porous zone as shown in \Cref{fig_schematic of permeable tip}, allows low-speed flows going through the permeable zone driven by the pressure difference between the \ac{ps} and the \ac{ss}, and these flows will further interfere with the \ac{ptv}. This is similar to the shrinking groove design (\cite{liu2018method}), but without the risks of new cavitation inside the groove. It also allows the numerical examination of the control effect across a wide range of permeability.

Existing studies by \cite{khorrami2002novel} and \cite{palleja2022reduction} for controlling tip leakage vortices using porous media focus specifically on the influence of porosity on the tip leakage noises in turbomachinery. A recent experimental and numerical work by \cite{bi2024numerical} on tip leakage cavitation around a foil demonstrates the promising potential of using the porous tip to mitigate cavitation, but they focused on tip leakage vortices across a range of tip gap sizes, and the role of permeability was not discussed. Furthermore, recent studies on porous coating around cylinders by \citeauthor{arcondoulis2023internal} and finlet rails on wing trailing-edge by \citeauthor{fiscaletti2024finlet} demonstrated that permeable structures have the potential to significantly mitigate vortex-induced noise.

In the present work, we propose controlling tip vortices, particularly the pressure-drop at the vortex core, through a confined permeable treatment. This approach is not limited to new blade designs utilising porous materials but can also be applied to structural modifications, such as incorporating multiple holes or grooves, to introduce a certain degree of tip permeability. By allowing controlled permeability at the blade tips, these designs aim to mitigate the intensity of tip vortices and reduce the associated risks of cavitation, wake, and noise. We apply a steady blade-resolved \ac{rans} approach to model a horizontal-axis tidal turbine and compare it to the existing data from experimental measurements (\cite{willden2023tidal}), as described in \Cref{sec:V&V} and \Cref{water tunnel experiment}. The tip permeability is modelled by changing the momentum equation in terms of adding the flow resistance in porous cell zone, and a wide range of permeablity is applied and discussed. In \Cref{sec:vortex pattern} and \Cref{sec: Cp at the vortex cores}, the vortex structures, minimal pressure-coefficient at the vortex core are compared and discussed to conclude the optimal permeability for mitigating cavitation risks. In \Cref{sec:CP & CT}, the influence of permeable tip treatment on the turbine power and thrust coefficients is also discussed. Finally, an investigation into the vortex parameters is performed to reveal the underlying physics.

\section{Case Configuration and Methodology} \label{sec:method}

\subsection{Turbine rotor simulation} \label{sec:turbine simulation method}
\begin{figure}[htbp]
	\centering
	\includegraphics[scale=.5]{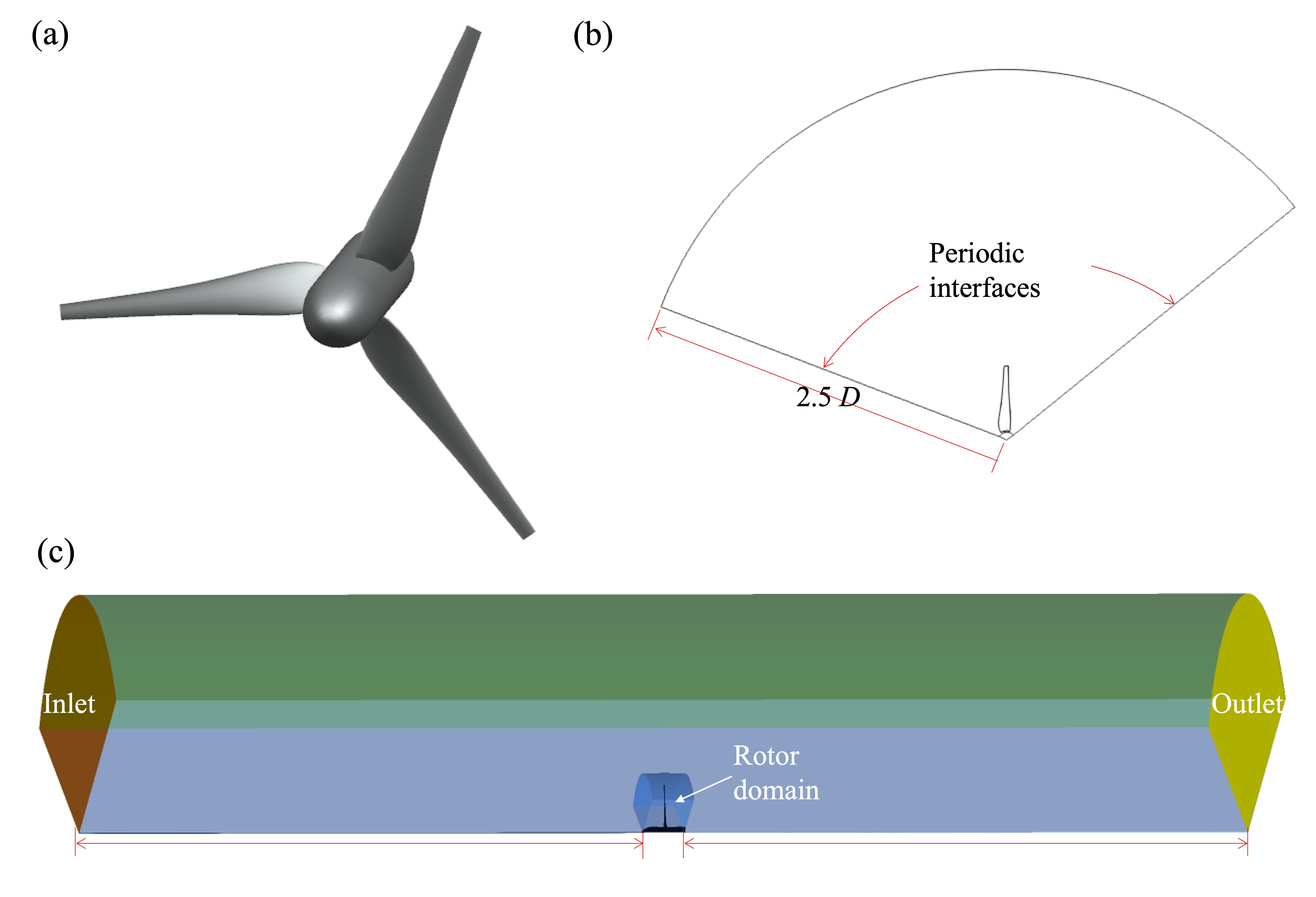}
    \caption{Modelled turbine rotor: (a) the complete geometry; (b) axial view of the 120$^\circ$-wedge computational domain. (c) three-dimensional view of the computational domain.}
\label{fig_model}
\end{figure}

In the present work, we model a third of a tidal turbine rotor (Figure~\ref{fig_model}) under a non-inertial rotating frame of reference using \ac{cfd} simulations. The model-scale turbine, tested by the UK Supergen Offshore Renewable Energy Hub (\cite{willden2023tidal}), has a rotor diameter of 1.6 m. The experiments were performed in a towing tank at a Reynolds number of $1.3\times10^6$ based on the constant towing velocity $U_\infty=1$~m~s$^{-1}$ and the turbine diameter. The turbine blade span is 0.7 m, and the blade tip-chord length is 0.043 m.

We solve the steady-state Reynolds-averaged Navier-Stokes (RANS) equations and employ the SST\:$k-\omega$ turbulence model within \textit{Ansys Fluent 2023R1}. The computational domain is a 120$^\circ$ slice of a cylinder, whose axis coincides with that of the tested turbine. The domain has a radius of $2.5D$. A rotor--stator interface is placed between the rotor domain and the stationary computational domain. The distance between the inlet and the front of the rotor hub, as well as the distance between the rear of the rotor hub and the outlet, is set to \(6D\) in the streamwise direction. Uniform, constant free stream velocity with low turbulence is set at the upstream inlet boundary; constant zero pressure is set at the downstream outlet boundary; periodic condition applied at the two inner side boundaries to account for the adjacent passages; slip condition applied at the outer side boundary; and no-slip condition at the blade and hub. The {\tt SIMPLEC} scheme is used to decouple pressure and velocity in solving the governing equations of incompressible flow. Second-order discretisation schemes are applied for gradients. A second-order scheme is used for pressure, and second-order upwind schemes are applied to the momentum equations, turbulence kinetic energy, and specific dissipation rate.

\begin{figure}[htbp]
	\centering
	\includegraphics[scale=.3]{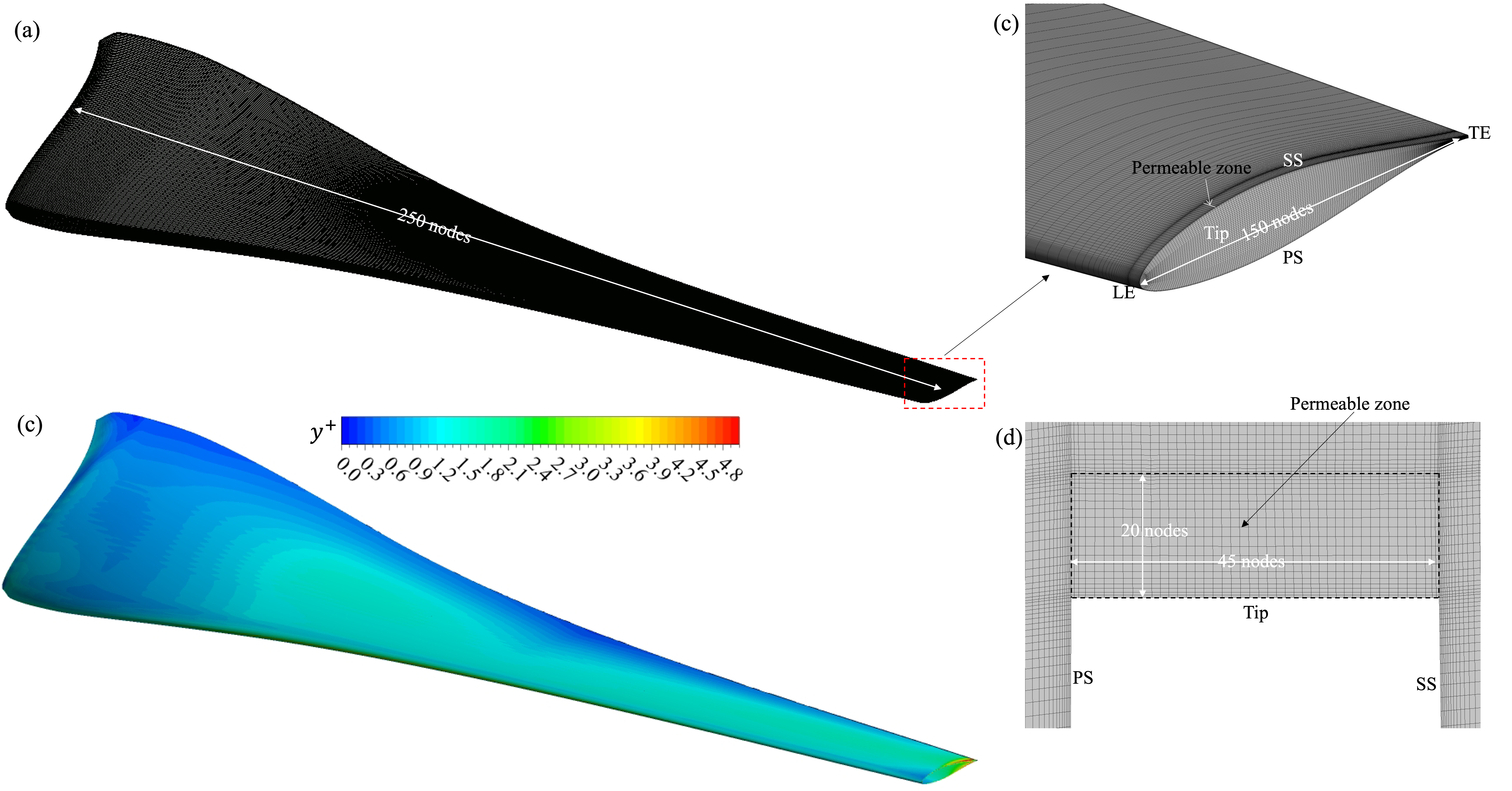}
\caption{Mesh distribution: (a) mesh on the blade surface; (b) refined mesh near the tip; (c) $y^+$ distribution on the blade surface; (d) mesh on a cut-section inside the permeable zone. LE: leading edge; TE: trailing edge; PS: pressure side; SS: suction side.}
\label{fig_mesh}
\end{figure}

ICEM-CFD is employed to discretise the computational domain with a structural mesh of around 17 millon hexahedral cells (Figure \Cref{fig_mesh}). The blade boundary layer is resolved with the averaged $y^+$ below 1 at the design tip-speed ratio $\lambda=6.03$, and the $y^+$ distribution on the blade surface is demonstrated in \Cref{fig_mesh}c. The growth ratio of the grid size inside the blade boundary layer is around 1.1 along the wall-normal direction. A total of 150 nodes are distributed along the blade tip in the chordwise direction, 45 nodes in the thickness direction, and 20 nodes in the spanwise direction within the permeable zone.

\subsection{Permeable zone simulation} \label{sec:permeability simulation method}
The permeable tip treatment is modelled by including a porous zone over the blade tip section. This porous zone is defined as a separate fluid zone in the computational domain and designated as a porous medium. As shown in \Cref{fig_mesh}d, the interface between the permeable tip and the surrounding flow field is treated using conformal mesh to ensure continuity in velocity and pressure fields.

To model the permeability in this zone, we used the built-in porous media model in \textit{Fluent}, where the momentum sink term is defined based on the Darcy model. In this model, the pressure loss due to the porous medium is proportional to the local velocity, representing viscous resistance only. The sink term added to the momentum equation takes the form:

\[
\boldsymbol{S} = -\frac{\mu}{\kappa} \boldsymbol{u}
\]

where \( \mu \) is the dynamic viscosity, $\kappa$ is the permeability, and \( \boldsymbol{u} \) is the velocity vector.

When this sink term is incorporated into the Navier--Stokes equations and subsequently non-dimensionalised, it results in the form of the following Darcy equation (\cite{neale1974practical,bose2023porous}): 

\begin{equation}
\frac{\partial \boldsymbol{u}}{\partial t}+(\boldsymbol{u} \cdot \nabla) \boldsymbol{u}=-\nabla p+\frac{1}{R e} \nabla^2 \boldsymbol{u}-\frac{1}{R e D a} \boldsymbol{u}
\label{eq_porous modelling}
\end{equation}

A uniform permeability $\kappa$ is considered throughout the porous zone, the effect of which is characterized by the non-dimensional parameter Darcy number ($Da = \kappa/\bar{\tau}^2 $), where $\tau$ is the tip thickness (shown in \Cref{fig_schematic of permeable tip}) and its averaged value is $0.26\%D$ along the chordwise direction.

This study is carried out for various $Da$ values in the range of $10^{-4}$ to $10^{-7}$, respectively. The spanwise extent of the porous zone is in the order of $0.1\%D$, and different spanwise extents have been investigated in the present work.

\section{Verification and Validation} \label{sec:V&V}

\subsection{Uncertainty assessment}
We assess the uncertainty following the least square approach proposed by \cite{viola2013uncertainty} and applied in our previous simulation work on a tidal turbine rotor (\cite{liu2023accuracy}).

\begin{figure}[htbp]
	\centering
	\includegraphics[scale=.6]{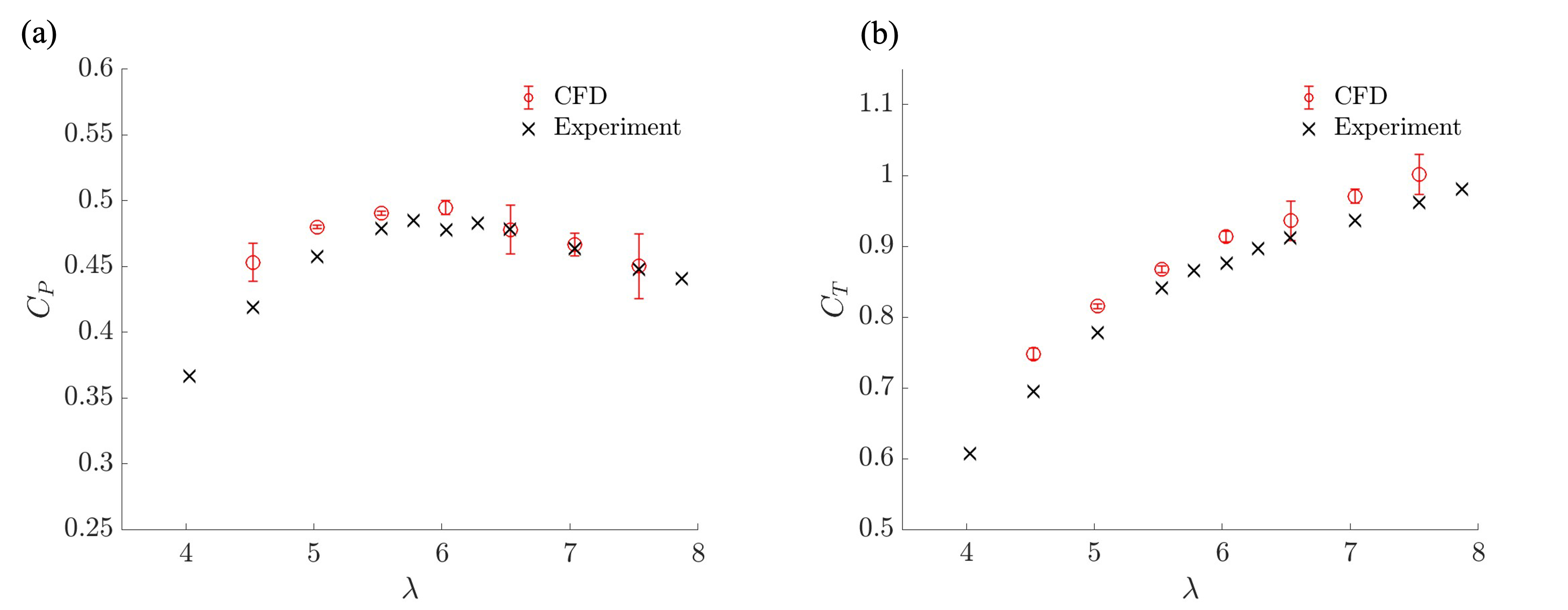}
    \caption{Comparison between CFD and experimental data (\cite{willden2023tidal}): (a) Power coefficient $C_P$ versus the tip-speed ratio $\lambda$; (b) Thrust coefficient $C_T$ versus the tip-speed ratio $\lambda$.}
    \label{fig_CPCT_Comparison}
\end{figure}

To quantify the reliability of the CFD predictions, we estimate the numerical uncertainty associated with each computed quantity of interest, denoted as \(\Phi\), representing either the power coefficient \(C_P\) or the thrust coefficient \(C_T\). The uncertainty is expressed in both absolute (\(E_\Phi\)) and relative (\(U_\Phi\)) forms at a 95\% confidence level, with the relationship defined as
\begin{equation}
E_\Phi = \Phi U_\Phi,
\end{equation}

The total relative numerical uncertainty \(U_\Phi\) accounts for several potential error sources, including discretization error due to finite grid resolution (\(U_{\Phi_g}\)), time stepping (\(U_{\Phi_t}\)), round-off error (\(U_{\Phi_r}\)), and convergence error from the iterative solver (\(U_{\Phi_c}\)). In this work, since the simulations are steady-state and use double precision, \(U_{\Phi_t} = 0\) and \(U_{\Phi_r}\) is considered negligible.

To evaluate the spatial discretisation uncertainty, simulations are performed using four sets of grid resolutions, spanning from $5.64\times10^6$ cells to $3.01\times10^7$ cells. A normalised metric is introduced as
\begin{equation}
\varphi(h) \equiv \frac{\Phi(h)}{\Phi(\mathrm{base})},
\end{equation}
where \(h\) represents the relative grid size and \(\Phi(\mathrm{base})\) is the solution obtained with the base mesh. The relationship \(\varphi(h)\) is approximated by a power-law fit of the form
\begin{equation}
\varphi(h) \approx \zeta h^\xi + \varphi_0,
\end{equation}
where the coefficients are either directly computed using the available data points or estimated via least-squares regression. The standard deviation \(\sigma\) of the fit quantifies the fitting error.

Depending on the fit quality \(p\), the relative uncertainty due to grid resolution is calculated as:
\begin{equation}
U_{\Phi_g} =
\begin{cases}
1.25 \left| 1 - \varphi_0 \right| + \sigma, & \text{if } p \geq 0.95, \\
1.5 \dfrac{\varphi_{\mathrm{max}} - \varphi_{\mathrm{min}}}{1 - \frac{h_{\mathrm{min}}}{h_{\mathrm{max}}}} + \sigma, & \text{if } p < 0.95,
\end{cases}
\label{eq:Uncertainty}
\end{equation}
where \(\varphi_{\mathrm{max}}\) and \(\varphi_{\mathrm{min}}\) are the extrema of \(\varphi(h)\), and \(h_{\mathrm{max}}\), \(h_{\mathrm{min}}\) are the corresponding step sizes.

The convergence error is estimated by fitting the solution history over 4000 iterations (excluding the first 2000) with an asymptotic function. The error is defined as the difference between the final solution and the asymptotic limit, scaled by a safety factor of 1.25, and augmented by the standard deviation of the fit.

The total relative numerical uncertainty is computed as:
\begin{equation}
U_\Phi = \sqrt{U_{\Phi_g}^2 + U_{\Phi_t}^2 + U_{\Phi_r}^2} + U_{\Phi_c},
\end{equation}
noting that \(U_{\Phi_c}\) is not included under the square root because it is not considered to be independent of the other terms.

Simulations were carried out for a range of tip-speed ratios from 4.52 to 7.54. \Cref{fig_CPCT_Comparison} shows $C_P$ and  $C_T$  computed with CFD and the experimental data (\cite{willden2023tidal}) and the numerical uncertainty of CFD. The relative error in the CFD predictions of $C_P$ is 1.1\% at the design tip speed ratio $\lambda=6.03$, while it is higher both at low and high tip speed ratios, shown in \Cref{fig_CPCT_Comparison}a. At low tip-speed ratios, the error is up to 3.2\% at $\lambda=4.52$, and it might be attributed to underpredicted flow separation. The overall numerical uncertainty is between 0.28\% to 5.4\% for $C_P$, including the grid uncertainty $U_{\Phi_g}$ between 0.28\% to 5.4\% and the convergence uncertainty $U_{\Phi_c}$ between 0.01\% and 0.52\%. As shown in \Cref{fig_CPCT_Comparison}b, the relative error of the CFD in predicting $C_T$ is always within 3.0\%, taking account into the grid uncertainty $U_{\Phi_g}$ between 0.4\% to 2.97\% and the convergence uncertainty $U_{\Phi_c}$ between 0.05\% and 0.3\% for $C_T$.

\subsection{Water tunnel experiment}

In addition, we have performed \ac{piv} experiments in a water tunnel on a half-wing model, which was extruded from the blade-tip profile of the aforementioned turbine. For further details about the experiment setup, please refer to \Cref{water tunnel experiment}.

\begin{figure}[htbp]
	\centering
	\includegraphics[scale=.6]{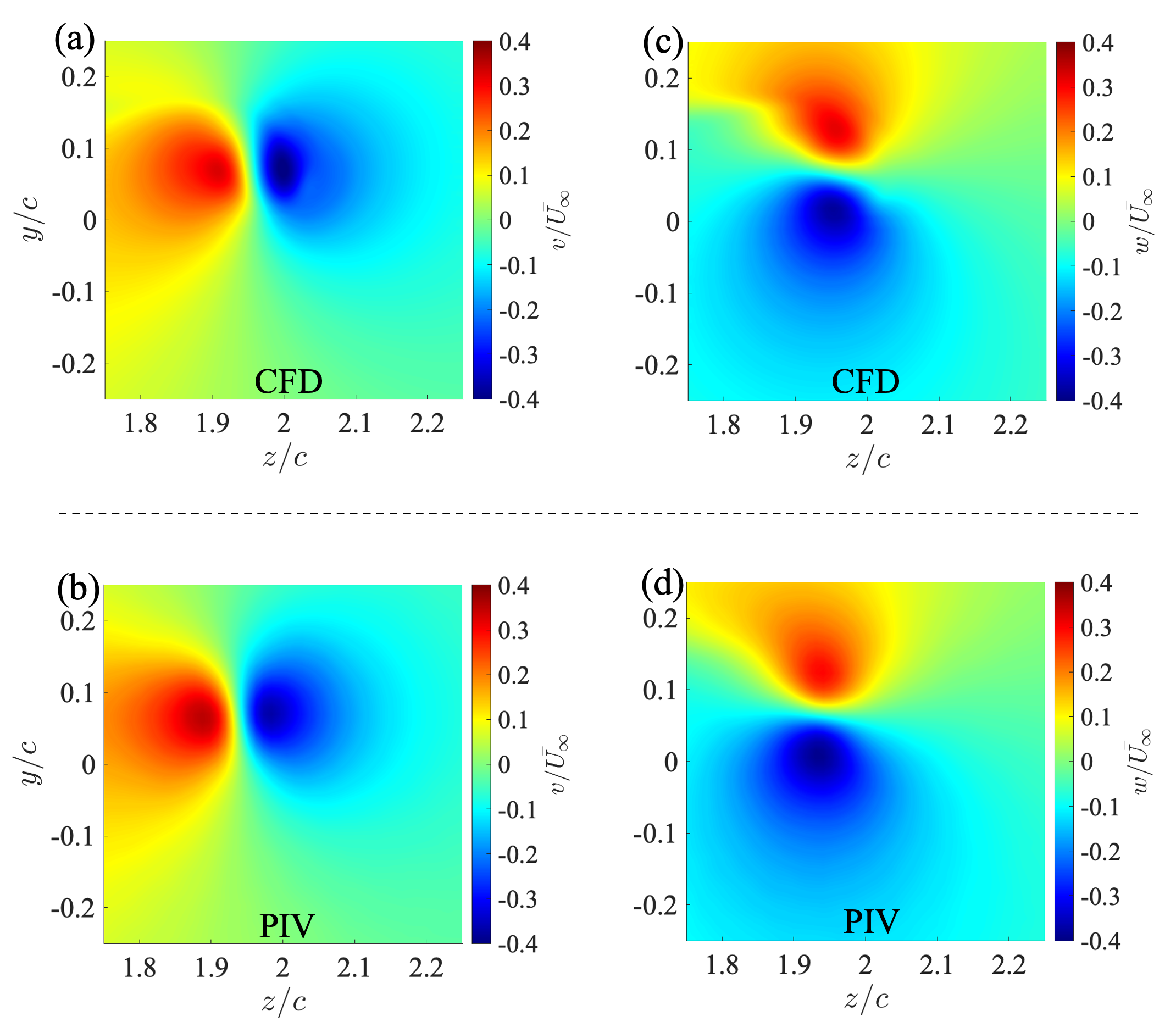}
\caption{Comparison of the pithwise velocity $v$ ((a), (b)) and spanwise velocity $w$ ((c), (d)) on the cross-section of $x/c=1.3$ between \ac{cfd} (((a), (c)) and \ac{piv} ((b), (d)). $U_\infty$ denotes the freestream velocity in the water tunnel.}
\label{fig_cfd & piv x/c=1.3}
\end{figure}

Numerical simulation on the half-wing model with the same configuration and flow conditions has been performed with the same CFD approach described in \Cref{sec:method}. The distribution of the pithwise velocity $v$ and spanwise velocity $w$ on the cross-section of $x/c=1.3$ is compared between the \ac{cfd} and \ac{piv}. \Cref{fig_cfd & piv x/c=1.3} shows good agreement between the simulation result and experiment result, which demonstrates that the employed \ac{cfd} approach can accurately predict the mean flow fields within the tip vortices.

\subsection{Permeability modelling validation}
Since no experimental data currently exist for permeable blade tip configurations, given that this is a novel concept introduced in our work, we have conducted a verification study for a porous (permeable) disk configuration against validated results available in existing studies (\cite{cummins2017effect, roos1971some}). The detailed description of the permeable disk case and simulation setup is described in \Cref{Permeable disk}.

\begin{figure}[htbp]
	\centering
	\includegraphics[scale=.2]{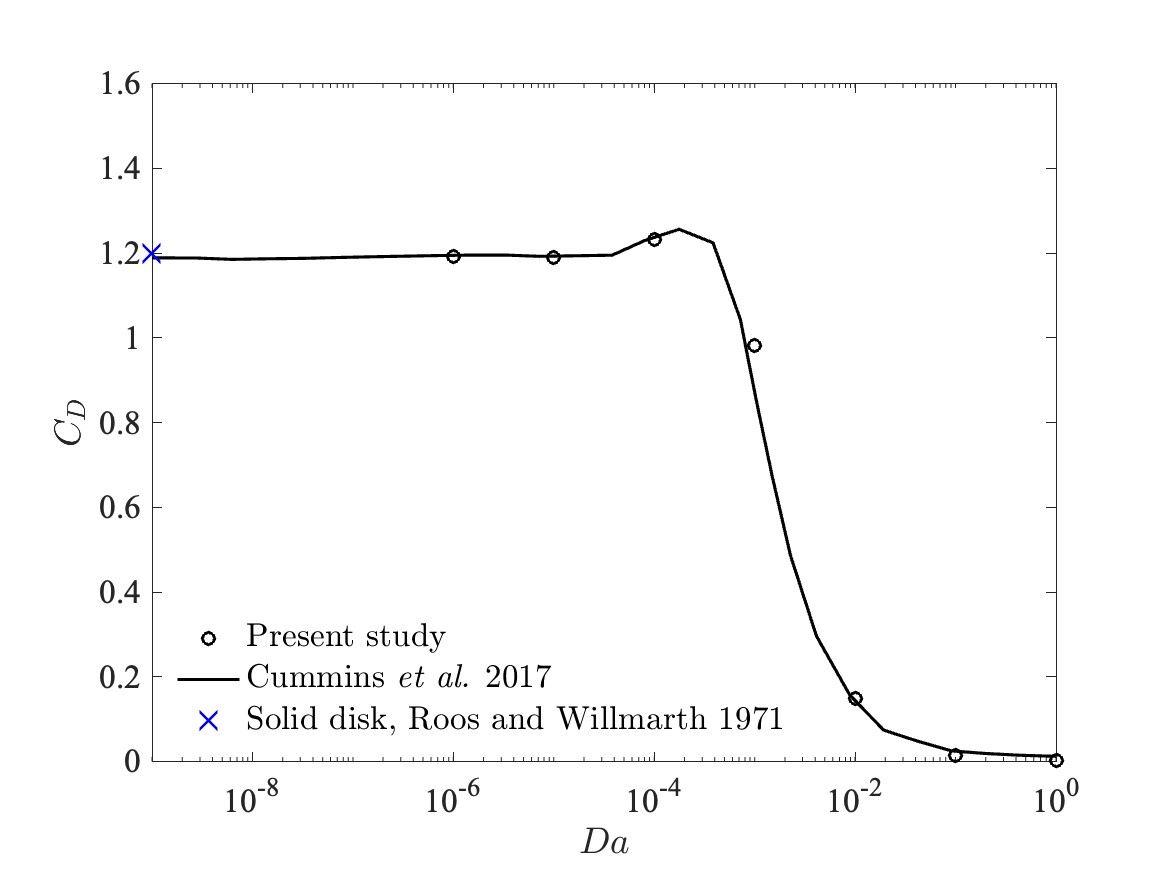}
    \caption{Drag coefficient $C_D$ versus Darcy number $Da$, in comparison with the results permeable disks from \cite{cummins2017effect} and impermeable (solid) disks from \cite{roos1971some}.}
    \label{fig_Permeable disk}
\end{figure}

The outcomes of this verification study are presented in \Cref{fig_Permeable disk}, which demonstrates good agreement between our numerical predictions and those reported in the literature for the predicted drag coefficients. When $Da \to 0$, the disk becomes effectively impermeable, and the predicted drag coefficient $C_D$ approaches the value corresponding to a solid disk. Conversely, as $Da \to \infty$, the predicted $C_D$ tends toward zero due to the absence of resistance in the permeable region. This trend is consistently observed in both our simulation results using the Darcy model and in existing literature.
This further supports the reliability of our implementation of the porous zone modelling approach in our present work.

\section{Results and Discussion}

\subsection{Vortex pattern} \label{sec:vortex pattern}

\begin{figure}[htbp]
	\centering
	\includegraphics[scale=.5]{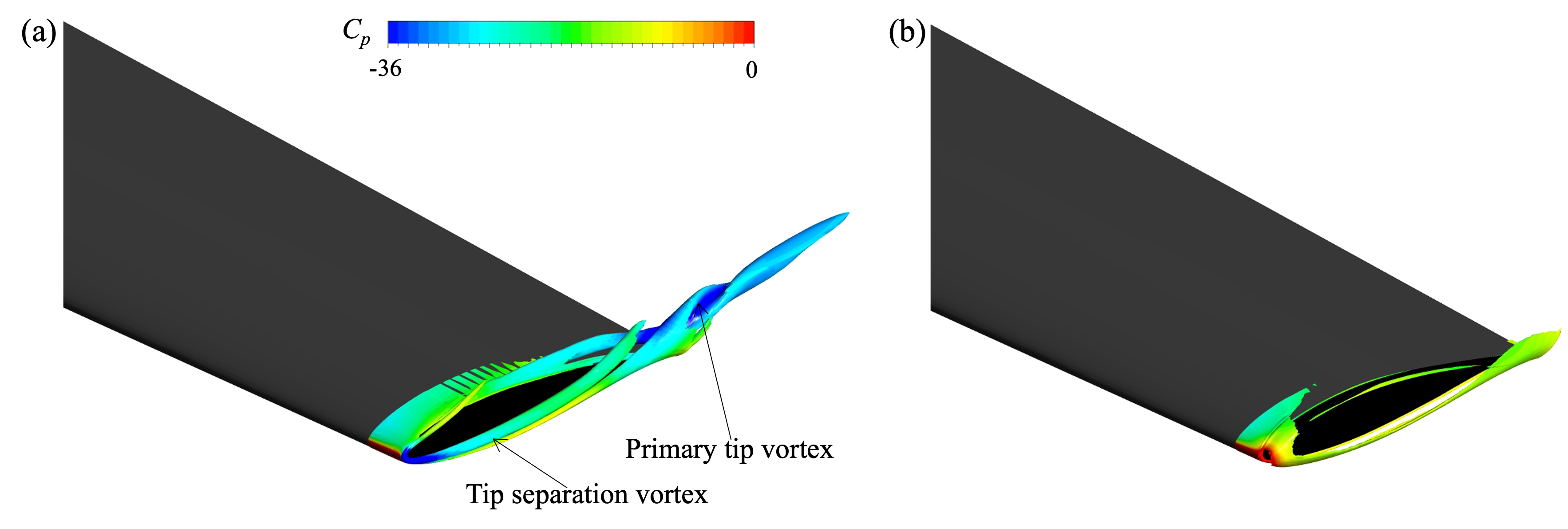}
\caption{Comparison of 3D vortex pattern between: (a) baseline (without tip permeability); (b) permeable tip with a non-dimensional permeability $Da=10^{-5}$. The vortex is visualised by the iso-surface of $Q D^2/U_\infty^2=7.68 \times 10^6$, where $Q$ is the second invariant of the velocity gradient tensor ($Q$ criterion). Pressure-coefficient $C_p$ is defined as $C_p=\frac{p-p_\infty}{\frac{1}{2} \rho U_\infty^2}$, where $p_\infty$ is the inlet pressure and $U_\infty$ is the towing velocity.}
\label{fig_IsoQ between baseline and Da1e-5}
\end{figure}

\begin{figure}[htbp]
	\centering
	\includegraphics[scale=.6]{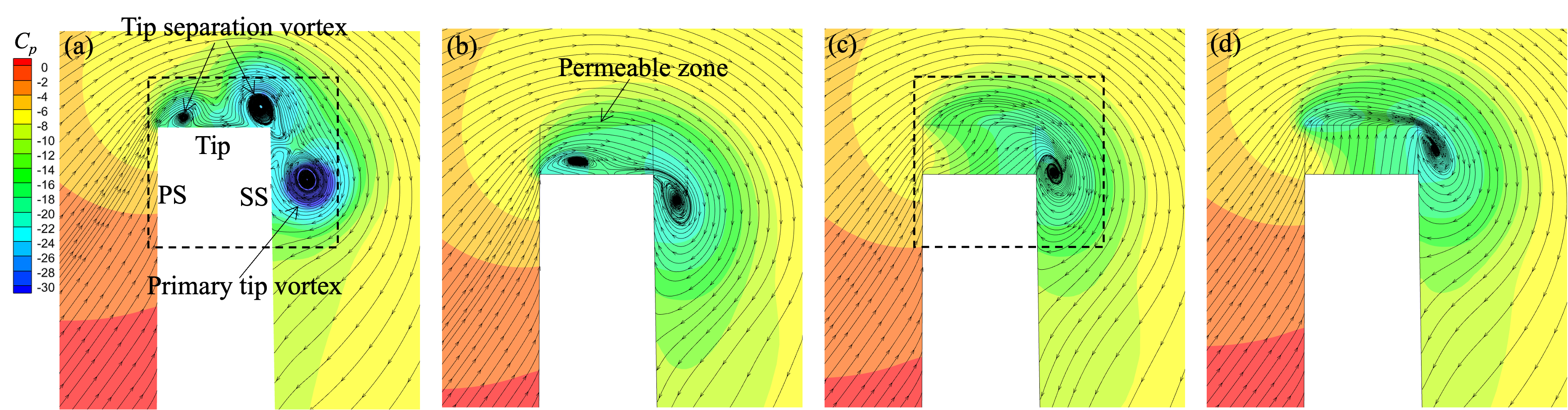}
\caption{Comparison of surface stream-traces (obtained by interpolating the velocity gradients) at the cross-section of 70\% chord between: (a) baseline; (b) permeable tip with a non-dimensional permeability $Da=10^{-4}$; (c) permeable tip with  $Da=10^{-5}$; (d) permeable tip with $Da=10^{-6}$.}
\label{fig_Surface stream-traces between baseline and permeable tip}
\end{figure}

\Cref{fig_IsoQ between baseline and Da1e-5} shows the comparison of tip vortices, including \ac{tsv} and \ac{ptv}, between the baseline case, which is the original turbine without tip permeability, and the permeable tip case, as described in \Cref{sec:method} and \Cref{fig_schematic of permeable tip} where $Da=10^{-5}$ and $\zeta=0.1\%D$. The permeable tip treatment leads to a prominent mitigation of \ac{ptv} and associated pressure-drop at the vortex core compared to the baseline case. Though the \ac{tsv} remains visible, the corresponding low-pressure distribution is diminished, caused by the reduced flow separation at the pressure-side edge of the blade tip as demonstrated in \Cref{fig_Surface stream-traces between baseline and permeable tip}.

\Cref{fig_Surface stream-traces between baseline and permeable tip} demonstrates the surface stream-traces, which are plotted in Tecplot by interpolating the velocity gradients, at the cross-section of 70\% chord. The flow fields crossing the permeable zone across a range of Darcy numbers between $10^{-4}$ and $10^{-6}$ are presented and discussed. Compared to the baseline case, a significant attenuation of both \ac{tsv} and \ac{ptv} is observed for the permeable tip cases, and the local pressure-drop near the vortex cores is suppressed. When the permeability is increased to $10^{-4}$, the flow separation near the tip-\ac{ps} corner again becomes remarkable. When the permeability is decreased to $10^{-6}$, a stronger \ac{ptv} is observed near the suction side. Therefore, there is an optimal range of tip permeablity that results in the most effective control of tip vortices.

\begin{figure}[htbp]
	\centering
	\includegraphics[scale=.38]{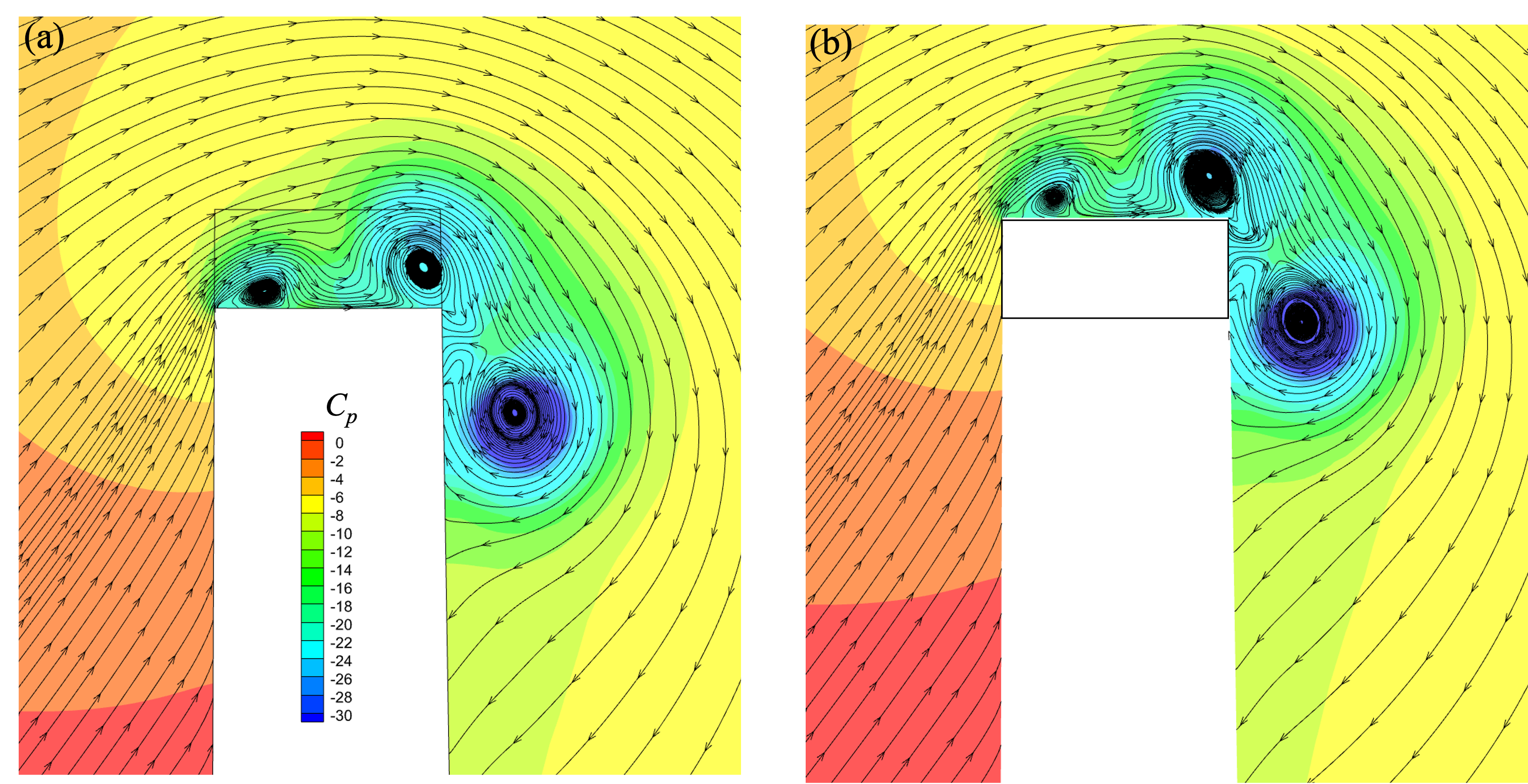}
\caption{Comparison of surface stream-traces at the cross-section of 70\% chord between: (a) $Da \rightarrow \infty$ (b) $Da \rightarrow 0$ (baseline).}
\label{fig_Surface stream-traces for Da_infty and Da_0}
\end{figure}

There are two extreme conditions of tip permeability we can theoretically consider: $Da \rightarrow \infty$ and $Da \rightarrow 0$. For $Da \rightarrow \infty$, the permeable zone presents no resistance to the fluids, and thus it is equivalent to the impermeable blade with a reduced blade span, i.e. the baseline blade with a shorter blade span. Therefore, the vortex pattern (\Cref{fig_Surface stream-traces for Da_infty and Da_0}a) becomes very close to that in the baseline case, For $Da \rightarrow 0$, the fluids cannot go through the zone at all, so the flow field recovers to the same as that around the baseline blade, shown in \Cref{fig_Surface stream-traces for Da_infty and Da_0}b.

\begin{figure}[htbp]
	\centering
	\includegraphics[scale=.45]{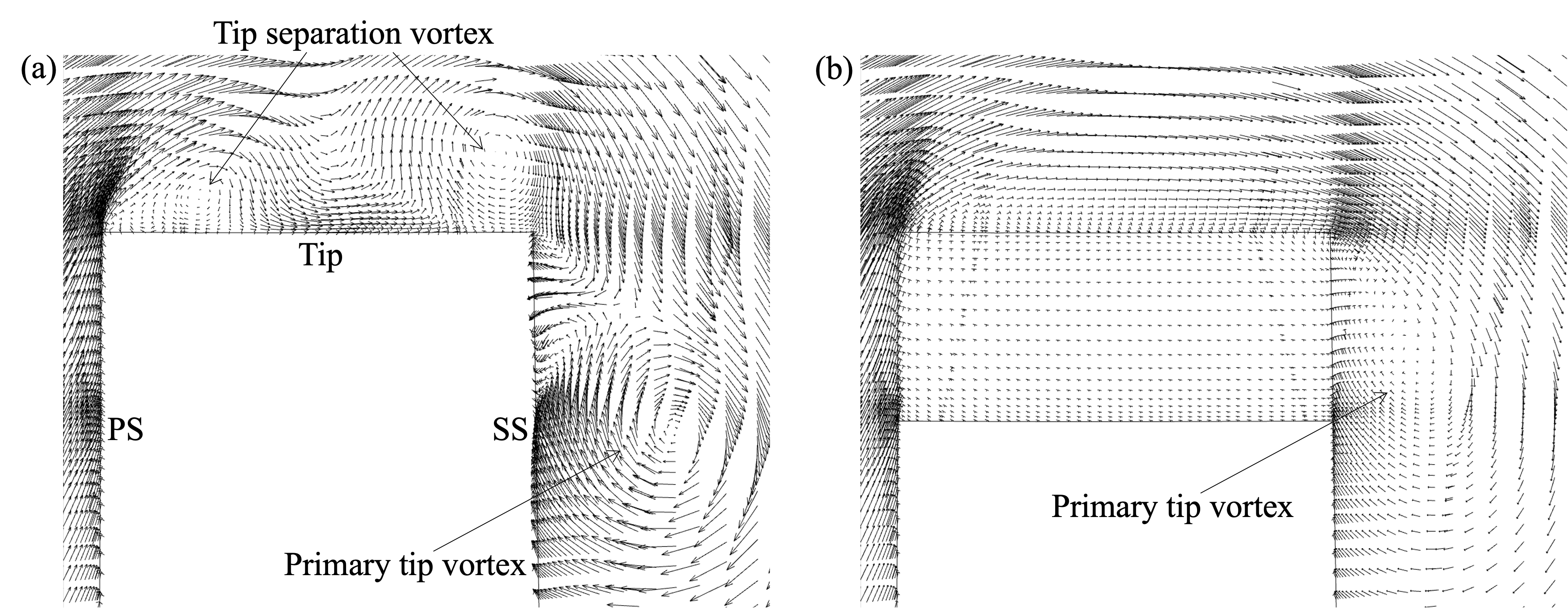}
\caption{Comparison of surface vectors at the cross-section of 70\% chord between: (a) baseline; (b) permeable tip with $Da=10^{-5}$.}
\label{fig_Surface vectors between baseline and Da1e-5}
\end{figure}

The surface vectors within the dashed-black box on the same cross-section in \Cref{fig_Surface stream-traces between baseline and permeable tip} for the baseline and the permeable tip with $Da=10^{-5}$ are shown in \Cref{fig_Surface vectors between baseline and Da1e-5}. Compared to \Cref{fig_Surface vectors between baseline and Da1e-5}a, less flow separation is observed at both the pressure-side and suction side around the tip when the permeable tip treatment is applied. A reduced swirling velocity around the \ac{ptv} core for the permeable tip (\Cref{fig_Surface vectors between baseline and Da1e-5}b) is also found referring to the vector field around the baseline blade. The above discussion suggests that tip permeability contributes to controlling tip vortices by suppressing the flow separation and vortex intensity near the tip.

\subsection{Pressure-coefficient at the vortex cores} \label{sec: Cp at the vortex cores}

\begin{figure}[htbp]
	\centering
	\includegraphics[scale=.5]{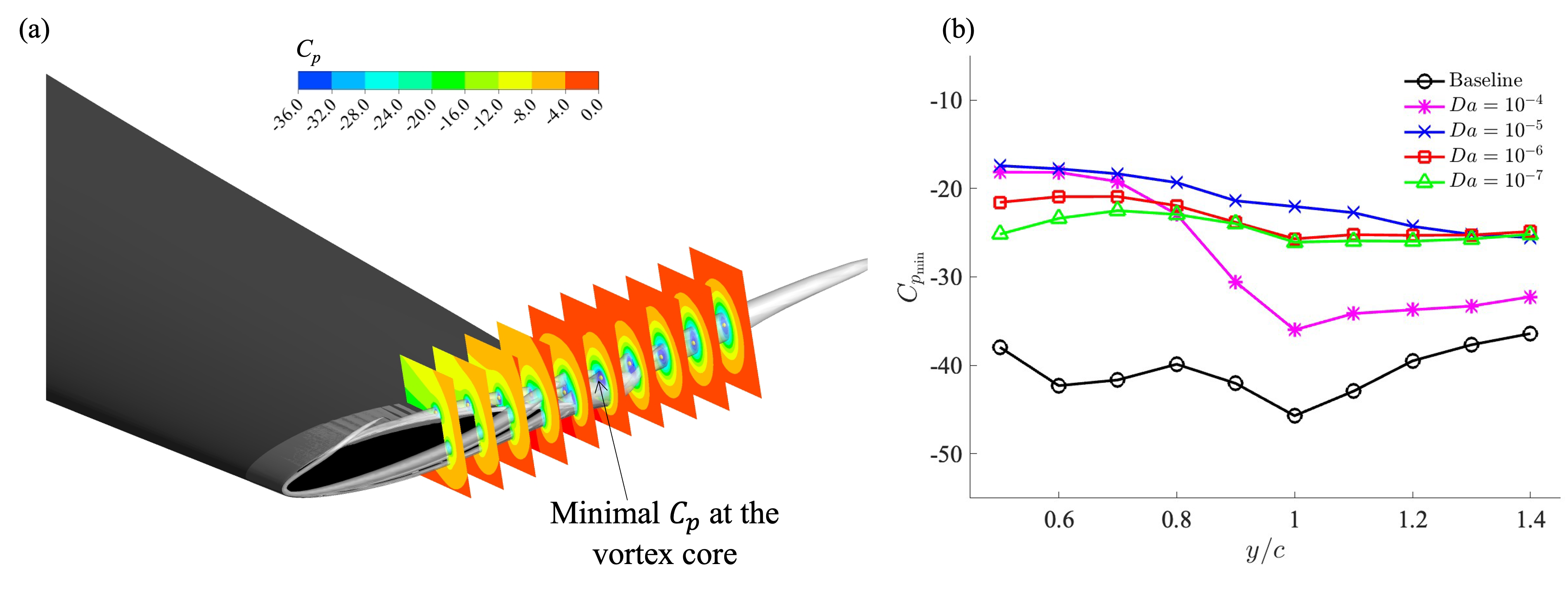}
\caption{Effect of tip permeability on mitigating pressure-drop: (a) tip vortices visualised by $Q$ criterion and distribution of pressure-coefficient $C_p$ on cross-sections across a range of chord ratios (b) comparison of minimal pressure-coefficient $C_{p_{\text {min}}}$ along the \ac{ptv} core trajectory between the baseline blade and the blade with different tip permeability. The tip speed ratio is 6.03.} 
\label{fig_Cp at vortex cores}
\end{figure}

As cavitation is our primary concern in the present work and the most significant pressure-drop occurs at the \ac{ptv} cores, we focus on the pressure-coefficient $C_p$ along the \ac{ptv} core trajectory. As illustrated in \Cref{fig_Cp at vortex cores}a, we extract the minimal $C_p$ on each cross-section between the mid-chord position of $y/c=0.5$, where \ac{ptv} just detached from the blade tip, to $y/c=1.4$ in the wake region. 

We further plot the distribution of the extracted minimal pressure-coefficent $C_{p_{\text {min}}}$ along the \ac{ptv} core trajectory (\Cref{fig_Cp at vortex cores}b). It is demonstrated that there is an optimal range of permeability, corresponding to $Da$ around $10^{-5}$, that can substantially reduce the tip vortex intensity and associated pressure-drop. This finding aligns well with the discussion on the vortex pattern with different tip permeability in \Cref{sec:vortex pattern}. The lowest value of $C_{p_{\text {min}}}$ along the \ac{ptv} core trajectory is found to be -45.7, and the permeable tip with $Da=10^{-5}$ increases it to -22.0, which means a relative change of 52\%. 

\begin{figure}[htbp]
	\centering
	\includegraphics[scale=.6]{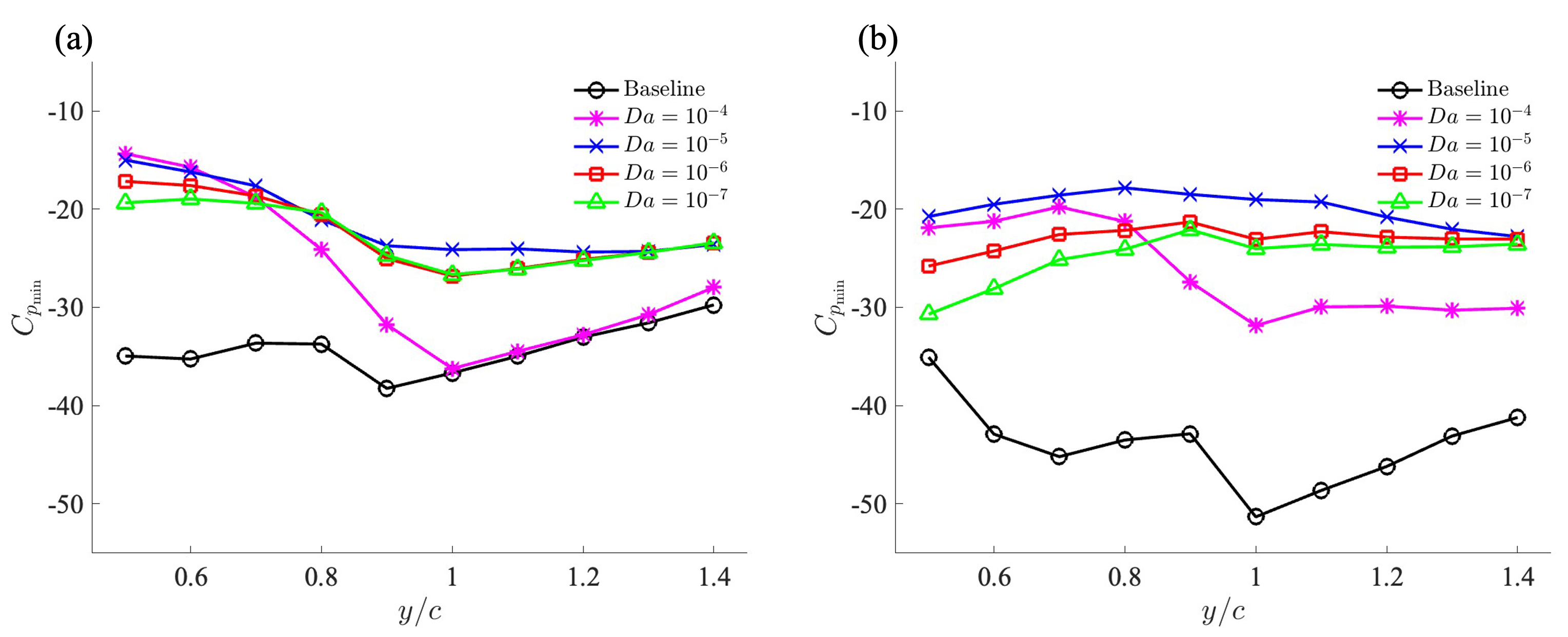}
\caption{Comparison of minimal pressure-coefficient  $C_{p_{\text {min}}}$ along the \ac{ptv} core between the baseline and the permeable tip at the tip speed ratio of: (a) $\lambda=4.52$ (b) $\lambda=7.54$.}
\label{fig_Cp at vortex cores under different tip speed ratios}
\end{figure}

A further investigation into the effects of tip permeability across different \ac{tsr}s and different spanwise extent of the permeable zone is presented in \Cref{fig_Cp at vortex cores under different tip speed ratios}. Between $\lambda=4.52$ to $\lambda=7.54$, all the results show a remarkable mitigation of the pressure-drop at the \ac{ptv} cores, and the permeable tip treatment with $Da=10^{-5}$ always show the optimal effect. Even promisingly, the mitigation effect becomes more significant at higher \ac{tsr}s, and the relative increase of $C_{p_{\text {min}}}$ reaches 63\%. 

These results demonstrate the great promise of using tip permeability to mitigate tip vortices and their cavitation risks, which will significantly contribute to breaking the current \ac{tsr} limit capped by blade-tip cavitation.

In addition, the influence of the permeable zone's spanwise extent is also considered and discussed in \Cref{appendix: spanwise extent}. The results suggest a more prominent mitigation of the pressure-drop with a larger spanwise extent of $\zeta=0.2\%D$. However, it also leads to a further drop in the power-coefficient while little change is observed for the thrust-coefficient. Therefore, there is a trade-off between mitigating tip vortices and ensuring a positive or negligible influence on the turbine's power and thrust performance.

\subsection{Power and thrust performance} \label{sec:CP & CT}

\begin{figure}[htbp]
	\centering
	\includegraphics[scale=.6]{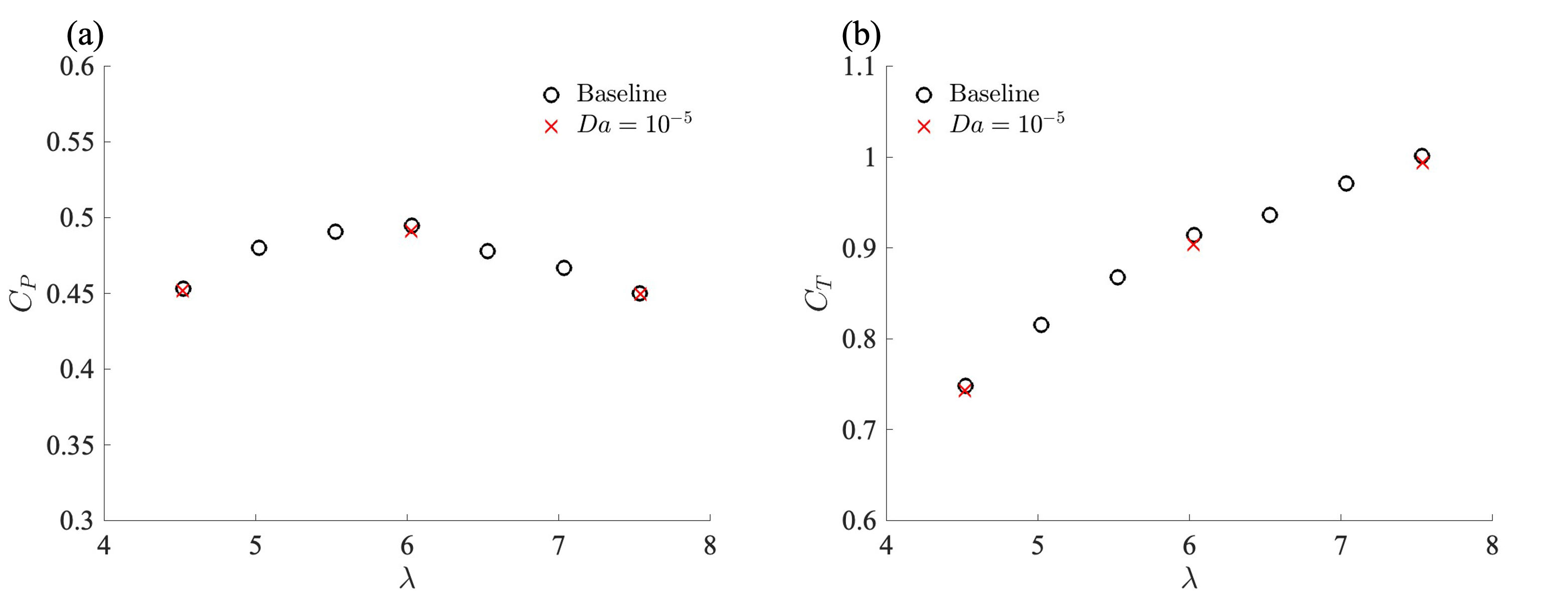}
\caption{Comparison of (a) power-coefficient $C_P$ and (b) thrust-coefficient $C_T$ between baseline and permeable tip with $Da=10^{-5}$ across a range of tip speed ratio $\lambda$ between 4.52 and 7.54.}
\label{fig_CP & CT between baseline and permeable tip}
\end{figure}

In this section, we discuss the influence of permeable tip treatment on the turbine's power and thrust performance. Simulations with and without the permeable tip show that the power-coefficient $C_P$ of the turbine with a permeable tip ($Da=10^{-5}$; $\zeta=0.1\%D$) drops by 0.25\% at $\lambda=4.52$, 0.75\% at $\lambda=6.03$, and 0.17\% at $\lambda=7.53$ compared to the baseline turbine (\Cref{fig_CP & CT between baseline and permeable tip}a). This suggests a negative but negligible influence on the turbine power-coefficient under non-cavitation conditions.

However, the thrust-coefficient $C_T$ decreases by 0.68\% at $\lambda=4.52$, 1.06\% at $\lambda=6.03$, and 0.72\% at $\lambda=7.53$ compared to the baseline turbine (\Cref{fig_CP & CT between baseline and permeable tip}b). A decreased $C_T$ is beneficial to reduce the levelised cost of energy for turbines, because it can allow reduced the structural loads, simplified powertrains and increased turbine lifespan. Moreover, the drop of $C_T$ is more significant than that of $C_P$. Therefore, the overall influence of tip permeability on the turbine's power-thrust performance is positive but limited.

\subsection{Vortex parameters and flow physics} \label{sec:vortex parameters}

To understand the underlying physics of how tip permeability reduces the intensity of tip vortices and associated pressure-drop, the vortex parameters are investigated in this section. Following the method applied by \cite{matthieu2015mind}, we analyse the minimum pressure-coefficient $C_{p_{\text {min}}}$ and the non-dimensional vortex intensity $\Gamma / (r_c U_\text{tip})$ at each cross-section, according to the following equation:

\begin{equation}
C_{p_{\text {min}}}=-\beta\left(\frac{\Gamma}{r_c U_\text{tip}}\right)^2
\end{equation}
\label{eq:Cpmin}
\noindent where $\Gamma$ is the circulation of the vortex; $U_\text{tip}= U_\infty \sqrt{1+\lambda^2}$ is the flow speed near the blade tip; $r_c$ is the viscous core radius where the maximal swirling velocity is reached; $\beta$ is a constant whose value depends on the choice of vortex model, and a Rankine vortex model is considered in the present work and $\beta=(2\pi^2)^{-1}$. In the present work, $\Gamma$ on each cross-section is computed by integrating the vorticity $\omega$ over an area where $\omega$ exceeds a specific threshold. As the boundary vorticity increases, the integrated circulation asymptotically approaches a limit. The threshold is selected when 99\% of the circulation is included.

\begin{figure}[htbp]
	\centering
	\includegraphics[scale=.25]{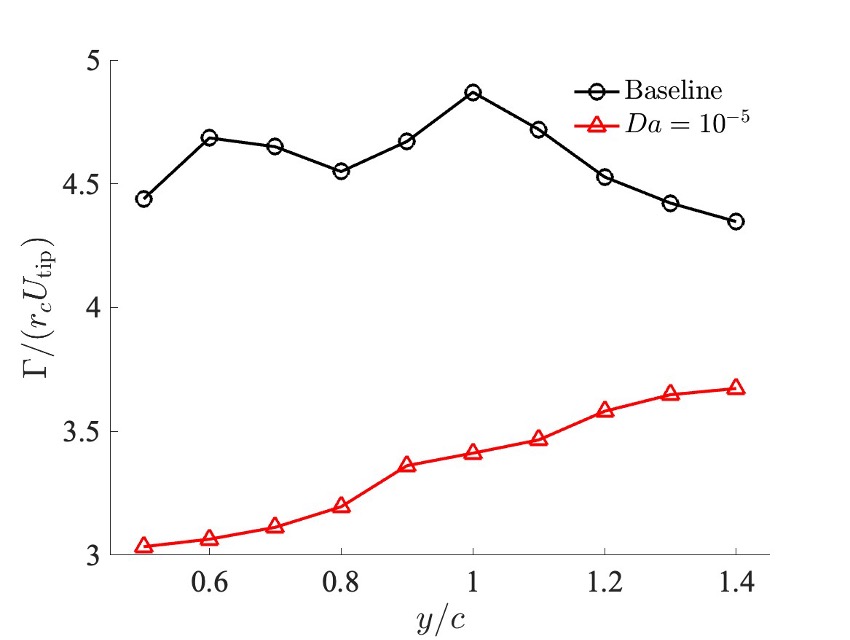}
\caption{Comparison of non-dimensional vortex intensity $\Gamma / (r_c U_\text{tip})$ on each cross-section along the vortex core trajectory between baseline and permeable tip with $Da=10^{-5}$. $r_c$ denotes vortex viscous core radius and $\Gamma$ denotes circulation. The tip speed ratio is 6.03.} 
\label{fig_non-dimensional vortex intensity}
\end{figure}

\begin{figure}[htbp]
	\centering
	\includegraphics[scale=.48]{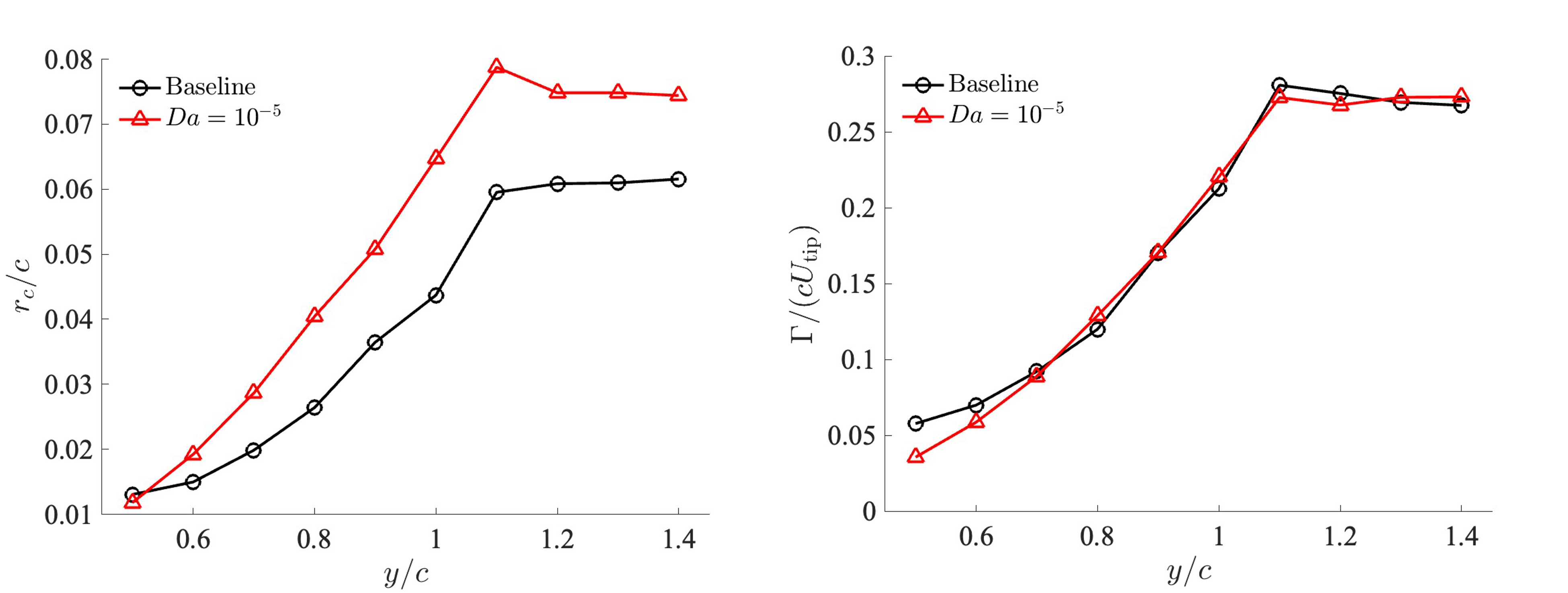}
\caption{Comparison of: (a) vortex viscous core radius $r_c$ (normalised by the blade-tip chord length $c$) (b) vortex circulation $\Gamma$ on each cross-section (normalised by $c$ and the flow speed $U_\text{tip}$ near the blade tip) along the vortex core trajectory between baseline and permeable tip with $Da=10^{-5}$. The tip speed ratio is 6.03.} 
\label{fig_radius and circulation}
\end{figure}

The non-dimensional vortex intensity $\Gamma / (r_c U_\text{tip})$ on each cross-section is extracted from the simulation and plotted along the \ac{ptv} core trajectory in \Cref{fig_non-dimensional vortex intensity}. The permeable tip case shows a significant drop in the non-dimensional vortex intensity compared to the baseline case, which corresponds to the reduction of $C_{p_{\text {min }}}$ in \Cref{fig_Cp at vortex cores}b.

\begin{figure}[htbp]
	\centering
	\includegraphics[scale=.5]{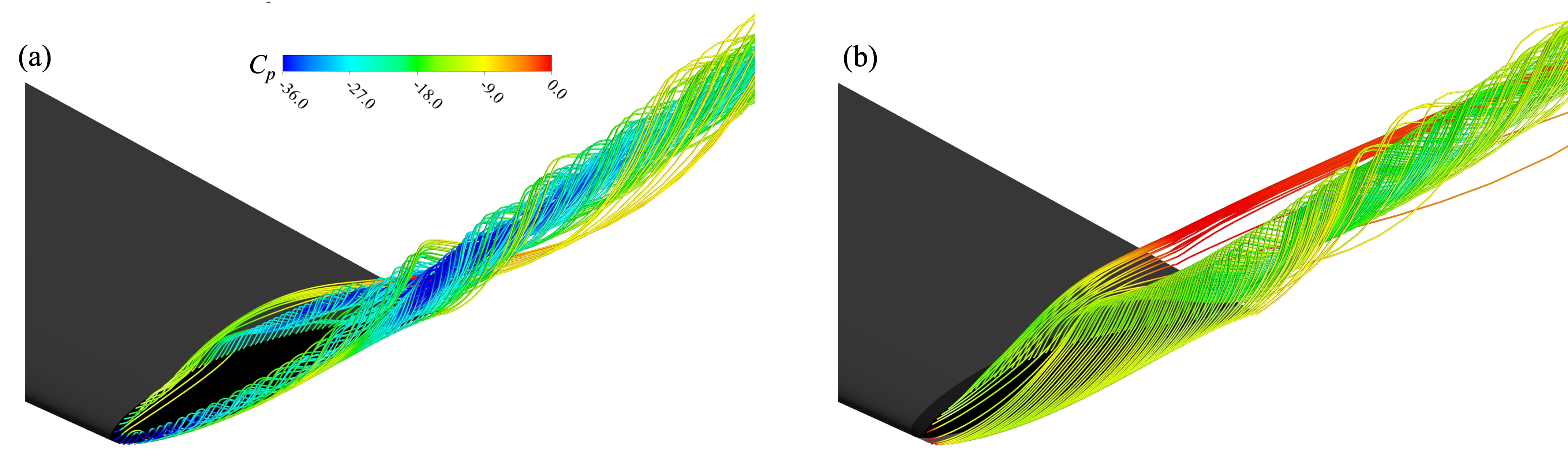}
\caption{Comparison of 3D streamlines released from the blade tip between: (a) baseline; (b) permeable tip with $Da=10^{-5}$ at a tip speed ratio of 6.03.} 
\label{fig_3D Streamlines}
\end{figure}

We further investigate the vortex radius $r_c$ and the circulation $\Gamma$ on each cross-section and plot them along the \ac{ptv} core trajectory, respectively. On the one hand, as shown in \Cref{fig_radius and circulation}a, the vortex radius is enlarged when the permeable tip is applied compared to the baseline case, which is also evident in the 3D streamlines released from the blade tip (\Cref{fig_3D Streamlines}). On the other hand, there is little change in the circulation of \ac{ptv} along the vortex core trajectory. These findings inform that the effective mitigation of vortex intensity through permeable tip treatment is mainly contributed by the enlarged radial dimension of the vortex. As illustrated in the comparison of 3D streamlines between the baseline (\Cref{fig_3D Streamlines}a) and the permeable tip (\Cref{fig_3D Streamlines}b), the application of a permeable tip results in a significant diffusion of the vortex over a larger area, thereby reducing its intensity. Consequently, the pressure-drop at the vortex core is mitigated.

\section{Conclusion and Future Work}

This work presents a novel passive, small-scale flow control strategy that introduces local permeability at the blade tip to mitigate trailing tip vortices and associated pressure drops, which can lead to cavitation. Unlike traditional methods that modify blade geometry or employ active control techniques, this study demonstrates that introducing a thin permeable zone—occupying only 0.1\% of the turbine diameter—can substantially alter the vortex dynamics while preserving the performance of blade-based systems.

A broad parametric analysis across tip-speed ratios (\ac{tsr} = 4.52–7.54) for a model-scale tidal turbine and Darcy numbers ($Da = 10^{-4}$–$10^{-7}$) reveals that optimal mitigation occurs at approximately $Da = 10^{-5}$. At this level of permeability, the tip vortex intensity is significantly reduced due to an enlarged vortex viscous core radius, while the circulation remains nearly unchanged. This leads to enhanced vortex diffusion, which in turn reduces the minimum suction pressure coefficient in the vortex core by up to 63\%, greatly mitigating cavitation risk.

The novelty of this work lies in identifying a new physical mechanism for vortex control—namely, the manipulation of tip vortex structure through local permeability without compromising the hydrodynamic performance of the blades. Furthermore, the consistent performance of the mitigation effect across a wide range of \ac{tsr}s indicates its robustness and applicability to real-world operating conditions. The minimal impact on power and thrust coefficients (less than 1\% in all cases) underscores the practical feasibility of this approach.

These findings provide a new direction for passive flow control via material and structural design. The concept of controlling tip vortices through permeability has broader implications for improving the aerodynamic and hydrodynamic performance of turbines, propellers, drones, and lifting surfaces in both underwater and aerial environments. 

A critical next step in this research is to deepen our understanding of the unsteady fluid dynamics, particularly the complex interactions among tip vortices, cavitation clouds, and the permeable tip structure. In particular, coupling high-fidelity approaches, such as \ac{les} with cavitation models, will be essential. The advanced cavitation modelling frameworks developed by \citeauthor{zhao2024insights} and \citeauthor{wang2023numerical} offer promising tools to capture these intricate dynamics with improved accuracy. 

Furthermore, future work will explore experimental validation and structural implementation strategies to replicate the permeable effect, starting from simplified foil models and progressing towards turbine blades. In addition, we aim to extend this concept to the control of wake dynamics and noise generation, with potential applications across a range of turbomachinery systems and the design of aerial and underwater vehicles.

However, while the concept of flow control with permeable structures is fundamentally novel and shows strong potential, its practical implementation may require high-precision manufacturing to realise fine-scale permeable structures. Furthermore, potential issues such as biofouling, local blockage, and long-term structural durability in harsh marine environments are important considerations. These challenges highlight the need for interdisciplinary collaboration across fluid dynamics, advanced manufacturing, structural mechanics, and materials science to advance this concept toward real-world application.

\newpage

\printcredits

\section*{Acknowledgments}
This work has been supported by the Royal Commission for the Exhibition of 1851 Brunel Fellowship (Yabin Liu), the UK EPSRC-funded Supergen Offshore Renewable Energy Hub [EP/Y016297/1], the Royal Society ISPF-International Collaboration Award [ICA/R1/231053], EPSRC Standard Proposal [EP/V009443/1], and the EPSRC IAA Innovation Competition [EPSRC IAA PV111].

\section*{Author Declarations}
\subsection*{Conflict of Interest}
The authors have no conflicts to disclose.

\section*{Data Availability}
Data will be made available on request.

\bibliographystyle{cas-model2-names}

\bibliography{cas-refs}

\newpage
\appendix

\section{Water tunnel experiments on a half-wing model} \label{water tunnel experiment}
\begin{figure}[htbp]
	\centering
	\includegraphics[scale=.4]{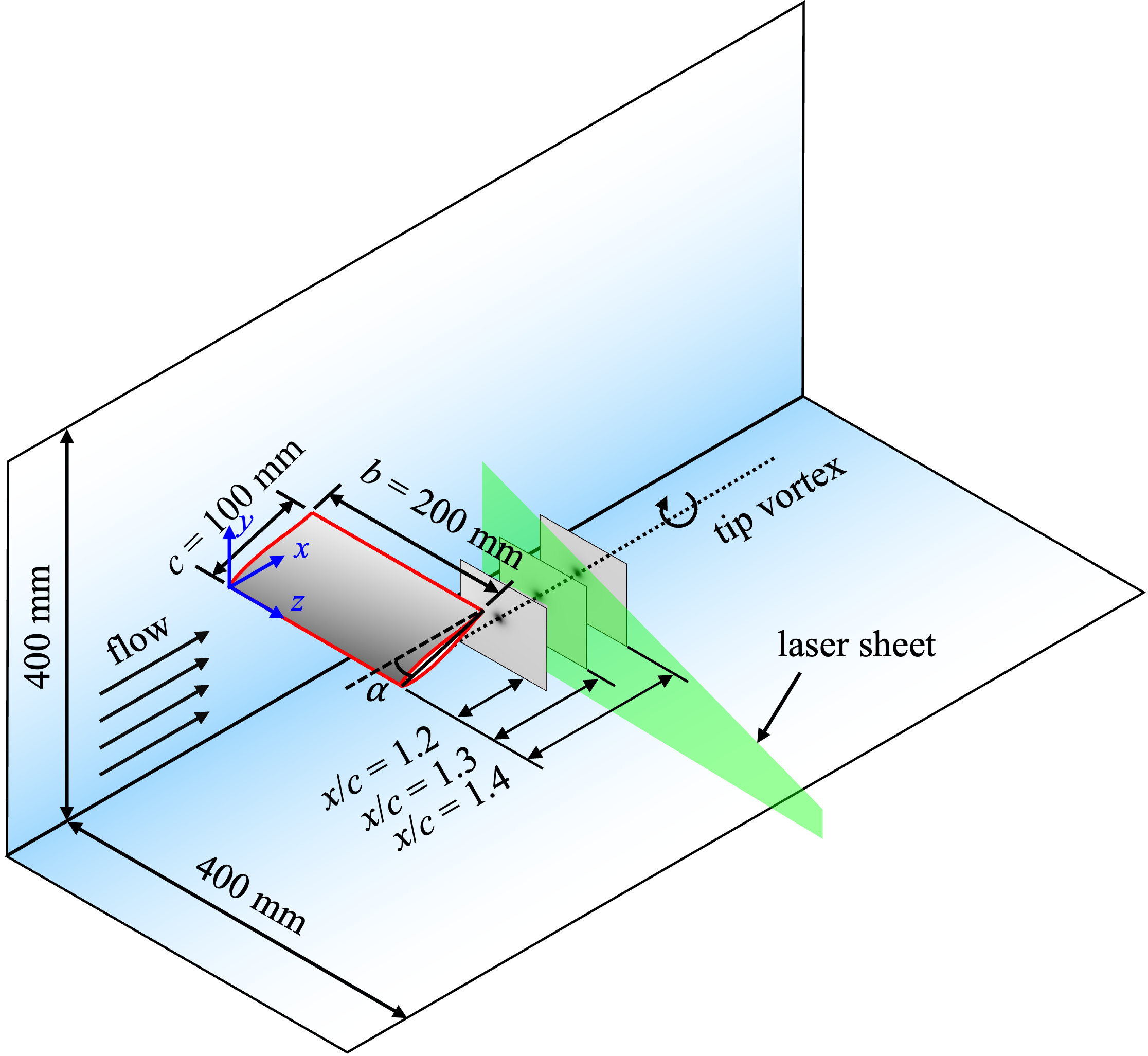}
\caption{Dimensions of the water tunnel and coordinate system used in the experiments. Displayed contours are obtained from CFD simulation and for illustrative purpose only.}
\label{fig_coordinate system for experiments}
\end{figure}

To validate the accuracy of the employed CFD approach on predicting tip vortices, \ac{piv} experiments were conducted in a water tunnel on a half-wing model, extruded from the blade-tip profile of the same turbine (\cite{willden2023tidal}). The water tunnel located at the University of Edinburgh is 8~m long, 0.4~m wide, and filled with water to 0.4~m depth, as shown in \Cref{fig_coordinate system for experiments}. The flow is preconditioned by curving vanes before entering and exiting the channel. The half-wing model has a chord length of $c=0.1$~m and a span of $b=0.2$~m. The coordinate system $x-y-z$ shows the streamwise, cross-stream, and spanwise coordinate. The flow velocity was calibrated using a Vectrino acoustic Doppler anemometer at the centre of the measurement plane without the presence of model and underwater cameras. The experiments were performed at a nominal free-stream velocity of $U_\infty=0.28$~m~s$^{-1}$ (equivalent to a Reynolds number based on the chord length of $Re_c=3\times10^4$) with a streamwise turbulence intensity of approximately 4.5\% before the leading edge. The wing's angle of attack is set as $6^ \circ$

\begin{figure}[htbp]
	\centering
	\includegraphics[scale=.45]{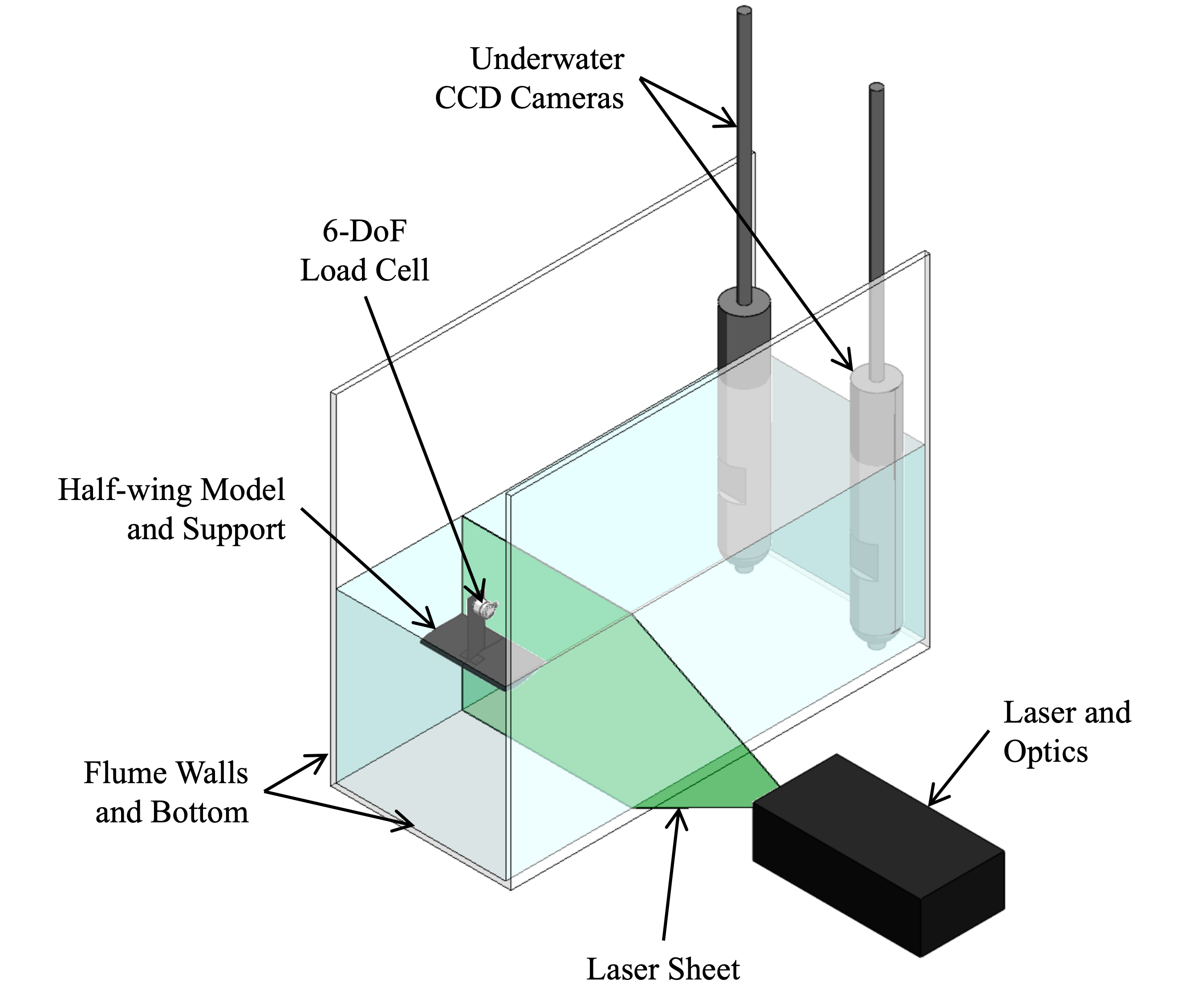}
\caption{Schematic of experimental setup for particle image velocimetry measurements for cross-flow planes.}
\label{fig_piv setup}
\end{figure}

The experimental setup for the \ac{piv} measurements and the tested sections are schematically shown in \Cref{fig_piv setup}. For more details, please refer to the work performed by \cite{muir2017underwater} with the same \ac{piv} and underwater camera system. Quantitative flow field analysis was performed using a LaVision FlowMaster Underwater stereoscopic \ac{piv} system, and the flow field cross-flow (normal to free-stream) sections at different streamwise locations were measured. For each measurement, 500 snapshots of the instantaneous flow field at each cross-section were utilised at a rate of 7~Hz to accurately account for the mean flow statistics. The interrogation window size was $32\times32$ pixels with 75\% overlap, producing an effective grid size of approximately 1\% of the chord length. The uncertainty of the velocity measurements was calculated to be within 2\% of the free-stream velocity.

\section{Permeable disk simulation} \label{Permeable disk}
\begin{figure}[htbp]
	\centering
	\includegraphics[scale=.4]{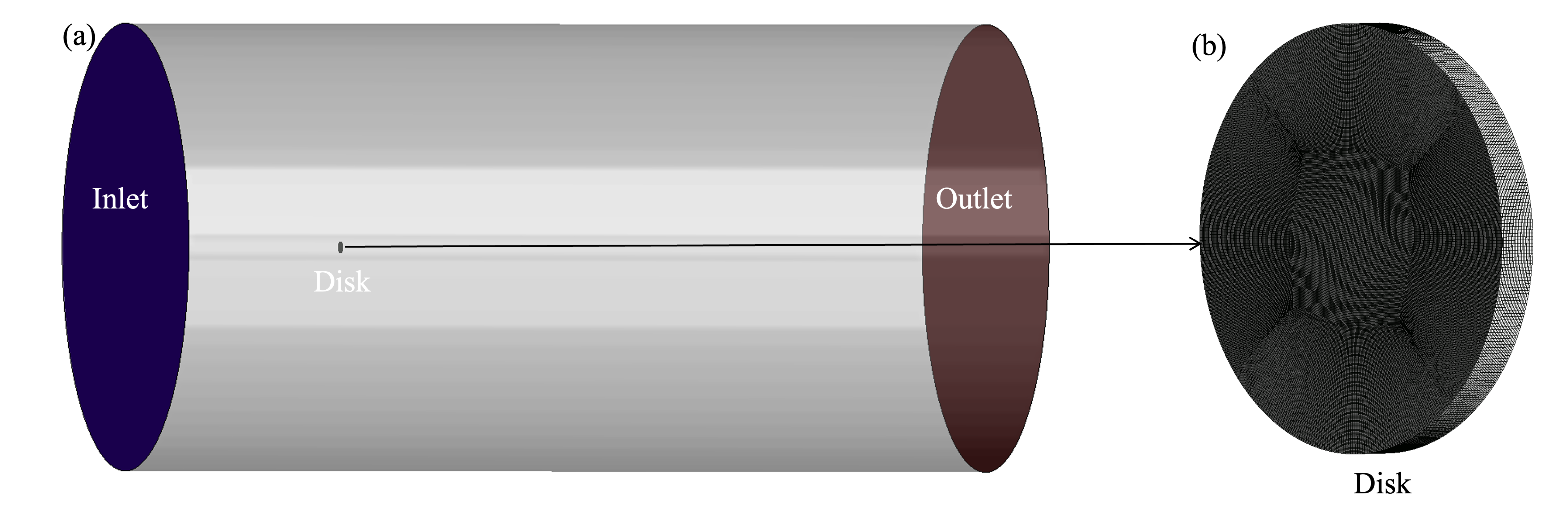}
\caption{Permeable disk case configuration: (a) computational domain; (b) mesh on the disk.}
\label{fig_Permeable disk case}
\end{figure}

To verify the permeability modelling approach, a validation study is conducted using a three-dimensional permeable disk, with reference to both numerical and experimental data from existing literature. As shown in \Cref{fig_Permeable disk case}, the disk has a thickness-to-diameter ratio of 10 and is positioned within a cylindrical computational domain that extends 60 disk diameters downstream and 40 disk diameters in the radial direction. The permeable disk is meshed with 100 cells radially, 400 circumferentially, and 20 across its thickness. The Reynolds number, defined using the freestream velocity and disk diameter, is set to $Re = 130$ based on available data from the literature (\cite{cummins2017effect, roos1971some}). Accordingly, a laminar flow model is applied in the permeable disk simulation.

The same porous media model, based on the Darcy equation and described in \Cref{sec:method}, is employed in this simulation. Drag coefficients of the permeable disk across a range of Darcy numbers have been extracted and compared with validated results from \cite{cummins2017effect, roos1971some}.

\section{Influence of the spanwise extent of the permeable zone} \label{appendix: spanwise extent}

\begin{figure}[htbp]
	\centering
	\includegraphics[scale=.5]{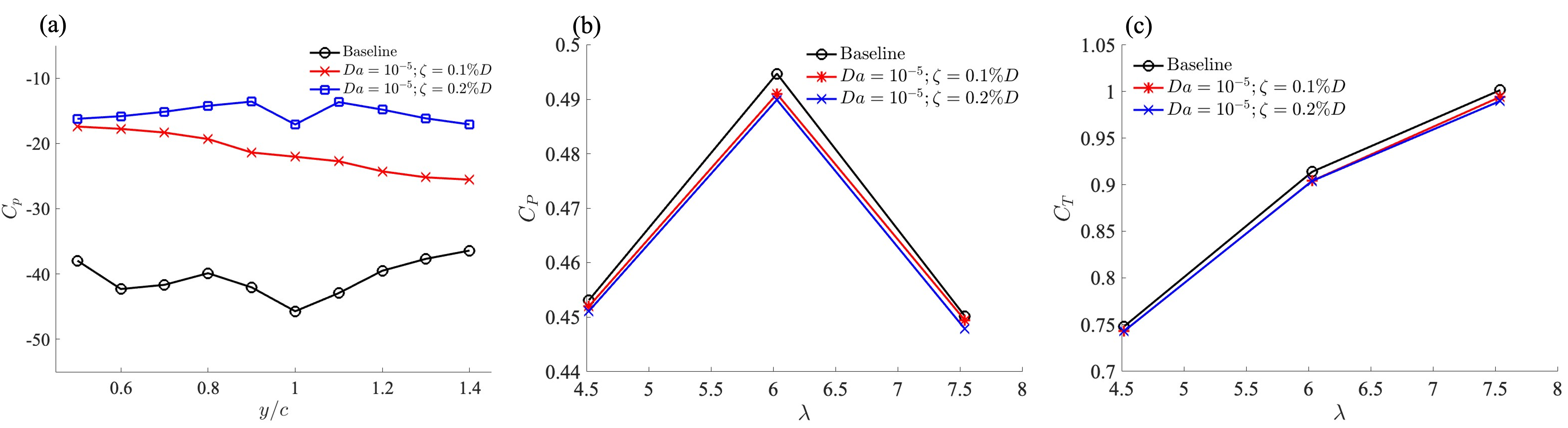}
\caption{Comparison of: (a) pressure-coefficient $C_p$ along the \ac{ptv} core (b) power-coefficient $C_P$ (c) thrust-coefficient $C_T$ between $\zeta=0.1\%D$ and $\zeta=0.2\%D$. The tip permeability is set as $Da=10^{-5}$.}
\label{fig_Cp CT and CT under different spanwise extents}
\end{figure}

The effects of the permeable tip with different spanwise extent of the permeable zone are investigated and discussed in this section. Though a more prominent mitigation of the pressure-drop with a larger spanwise extent of $\zeta=0.2\%D$ (\Cref{fig_Cp CT and CT under different spanwise extents}a), there is a further drop in the power-coefficient (\Cref{fig_Cp CT and CT under different spanwise extents}b), while little change is observed for thrust-coefficient (\Cref{fig_Cp CT and CT under different spanwise extents}c). Therefore, the negative influence on the power performance of the turbine is increased with a larger permeable zone. A trade-off between the effect of mitigating tip vortices and the turbine power-thrust performance should be considered in practical designs.

\end{document}